\documentstyle[amssymb,amsmath]{article}
\begin{document}

\def\slash#1{\setbox0=\hbox{$#1$}#1\hskip-\wd0\hbox to\wd0{\hss\sl/\/\hss}}

\centerline{{\Large\bf 
Compact Hyperbolic Manifolds as Internal Worlds}
\footnote{Talk presented at Cairo International 
Conference on High Energy Physics (9-14 January 2001),
Cairo, Egypt. Proceedings to be published by Rinton Press.
}}
\vskip 1truecm
\centerline{Rula Tabbash\footnote{e-mail: rula@sissa.it}}
\bigskip
\centerline{\it SISSA-ISAS, via Beirut 4, I-34014 Trieste, Italy 
}


\vskip 1.2truecm

\begin{abstract}
{
We comment briefly on some of the advantages and disadvantages
of compact hyperbolic manifolds 
as candidate manifolds for large radii compactifications.}
\end{abstract}


\section{Introduction and Description of The Problem}


The existence of extra dimensions, beyond the known four, 
seems to be a crucial ingredient in unifying gravity with 
the other gauge forces. 
The only promising quantum theory
for gravity, so far, is string theory and its mathematical 
consistency requires the space-time to have ten dimensions. 
The low energy field theoretical description of string theory
can be used in order to explain our four dimensional world
using the standard Kaluza--Klein compactification scenario, 
where the manifold of the $10$-dimensional space time, $W$, 
is a tensor product of our $4$-dimensional space-time, and 
an internal $6$-dimensional manifold. 
The conventional radius of compactification in string theory
is of order $M_{pl}^{-1}$, resulting in a $10$-dimensional 
gravity scale comparable to the $4$-dimensional Planck Mass, 
$M_{pl}\sim 10^{19}$GeV.

Inspired by the above theory, the last few years have witnessed
an increasing 
interest in adopting actions in more than four dimensions, however
performing the compactification on larger radii,
$R\gg M_{pl}^{-1}$.
The most interesting 
feature of such proposals 
\cite{dvali} is that it suggests a new way, 
different from 
grand unified theories
and supersymmetry,
to solve the hierarchy 
problem between the electroweak scale  
and the four dimensional gravity 
scale.\footnote{Other models with extra dimensions
using warp product have been proposed \cite{warp}, 
but here we limit our 
discussion to the standard tensor product case.}
Since the two scales 
are related by the volume of the $d$-dimensional internal space
$$
M_{pl}^2=R^d M^{d+2}
$$
As can be read from the above relation, the $4+d$ gravity scale 
can be lowered, e.g. down to TeV, by adjusting the radius 
$R$ appropriately. 

As a consequence, the solution of the hierarchy 
problem will be at the cost of introducing
new and unexplicabely small mass parameters to 
the effective four dimensional theory represented by the inverse 
radius of compactification, depending on how much one lowers
$M$. 

For $M\sim $TeV, this radius can be as big as 
$10^{13}$GeV$^{-1}$ for two extra dimensions, 
$10^{8}$GeV$^{-1}$ for three, $10^{5}$GeV$^{-1}$
for four, and so on. This problem becomes milder as 
the number of extra dimensions increases, however 
it may be desirable to avoid this shortcoming
even for a small number of them.
The primary problem is not explaining the smallness of these radii, 
which is obviously a fine-tuning problem, 
but rather avoiding their undesirable contributions to well 
studied observables. 
Upon compactifying down to four dimensions, one may 
in general get new degrees of freedom added to the 
spectrum of the Standard 
Model. The new states can be purely from the gravitational
sector, or have Standard Model Kaluza--Klein 
excitations in addition (depending
on
whether the SM interactions are written directly in four dimensions,
using the induced metric, or written fully in $D$ dimensions).
In any case, the new states, having masses of order $1/R$,  might 
lead to detectable modifications of 
the existing accelerator data and cosmological observations
\cite{dvali2}. This lead to  
imposing 
judicious bounds on the parameters of these theories. Whether these
bounds are
implemented
or not,
the theories with large extra 
dimensions experience difficulties
in realizing complementary scenarios
like the standard Cosmological one. \footnote{
For example, 
imposing an upper bound on the reheating temperature 
in order to avoid 
overproduction of Kaluza--Klein modes of the graviton, and 
discrete symmetries in order to prevent a fast proton decay makes
it difficult to construct a
baryogenesis model. Moreover, recovering 
the standard Friedmann-Robertson-Walker Universe in 
$4$ dimensions starting from higher dimensional 
Einstein's equations is not straightforward .}

Recently, it was suggested  
\cite{kn:m} that adopting a compact 
hyperbolic manifold (CHM) as the internal world may solve the 
fine-tuning problem in the radii of compactification,  
mentioned  
above,
utilizing only certain geometrical 
properties of the internal manifold, namely 
its exponentially large volume.
In addition, and
as a consequence of being able to choose the radius of compactification 
to be of the  order of the gravity scale,  
the undesirable contributions  
of the Kaluza--Klein degrees of freedom could be avoided
in a natural way (we will discuss this issue in seciotn
\ref{dis}).

Although 
hyperbolic spaces with finite volume are discussed in string and M-theory
\cite{Kehagias:2000dg}, 
their existence as a solution of the Einstein's equation should be 
examined, depending on the theory at hand and the choice of the 
metric in 
the $4$-dimensional 
manifold. 


\section{Compact Hyperbolic Manifolds}


A generic hyperbolic manifold, $H^d$ (such as the upper half
$d$-plane), has two 
main features of relevance to 
our discussion: 
\begin{itemize}
\item{ 
The volume depends exponentially on 
the curvature of the manifold.
}
\item{
Constant negative curvature.
 }
\end{itemize}  
The first feature was
pointed out in \cite{kn:m} and seems to offer a more statisfactory 
solution for the hierarchy problem than the ``classical''
scenarios with large extra dimensions. 
The second feature, on the other hand,  
is particularly attractive in our opinion
\cite{geo}, 
mainly because negatively curved manifolds admit harmonic spinors, 
a feature which is 
not yet well appreciated by physicist investigating models with 
large extra dimensions. 
This may be due to the fact 
that 
getting 
massless fermions on a compact manifold
does not mean 
achieving a chiral theory in $4$ dimensions.
In
fact, 
a
further 
coupling to a topologically non-trivial background 
will be generally needed in order to get rid of 
the extra spinorial degrees of freedom (for more
details see
\cite{sss1}).
In the following we will present in more details
how can each of those two properties help in 
solving {\it some} of the above mentioned problems.
Before we do so, it should be mentioned that 
a hyperbolic manifold, as it is well known, 
has an infinite volume. However, it is possible 
to compactify it, and hence to obtain a
CHM with finite volume, 
by moding out by an appropriate discrete subgroup, $\Gamma$, of 
its isometry group.\footnote{ In order to avoid
singularities, we chose $H^d/\Gamma$ with $\Gamma$ acting freely.}  
As an example, let us consider $H^2$ (the upper-half plane).
We can chose the following metric on it
$$
g_{ij}^{(d)}
dy^idy^j= \frac{1}{R^2}dr^2+ \mbox{sinh}^2(r/R)d\theta^2
$$
One can get a compact $H^2$ by moding out by the isometry group
$SL(2,Z)\subset SL(2,R)$. A sphere of radius $R$ cut out of $H^2$
will have the volume:
$$V=2\pi R^2[\mbox{cosh}(L/R)+2 L/R]$$
\noindent
Where $L$ is the diameter (largest distance) of $H^2$.
The relation between 
the volume and the curvature of the manifold is known 
as {\it rigidity}.
For $L>R$, one can write 
$$V=2\pi R^2\mbox{e}^{L/R}$$
We argue, as in \cite{kn:m}, that in the limit
$L> R$ the volume  
for a generic compact $H^d$ with diameter $L$
is:
\begin{equation}
V=a R^d\mbox{e}^{(d-1)L/R} 
\label{v}
\end{equation}
\noindent
where $a$ is a numerical factor, and the curvature
${|\cal R|}=R^{-2}$. The relation 
(\ref{v}) applies in general, with the exception of 
$d=3$, where the rigidity property of CHMs breaks down.


\section{Some Phenomenological Advantages of CHMs}


In the following we will consider theories of the form 
$$W=
M_4\times (H^d/\Gamma)$$ 
(as we mentioned previously, this ansatz
should be verified case wise). 
The metric will be
$$ds_W^2=g^{(4)}_{\mu\nu}(x)dx^\mu dx^\nu+R^2g_{ij}^{(d)}
(y)dy^idy^j$$
\noindent
Where 
$\mu,\nu=0,...3\;, \; i,j=1,...d$.\footnote{For 
other phenomenological and cosmological 
implications see \cite{kn:m,Starkman:2001dy}.}


\subsection{Approaching the Hierarchy Problem}


Let us start our discussion from the gravity sector 
(as it is the sector which determines the link between 
$M$ and $M_{p}$).
Consider some metric fluctuations and look at the 
linearized Einstein's equations: 
$
\Delta h_{MN}= \Delta_4h_{MN}(x,y)+\Delta_{H^d} h_{MN}(x,y)=0$.
Upon compactification, the metric fluctuations will fall into
various representations of the four dimensional reparameterization
group. Namely, 
$h_{\mu\nu}(x,y)$ is a graviton in $M_4$ and scalar in $H^d$;
$h_{i\mu}(x,y)$  is a vector in both $M_4$ and $H^d$; and 
$h_{ij}(x,y)$ is a scalar in $M_4$ and spin-2 field in $H^d$.
The mass spectrum of the various bosonic fields 
(including vector bosons) 
will be determind 
by the eigenvalues, $\lambda_n$ (to be identified by the 
mass$^2$ in the effective four dimensional theory), 
of the Laplacian on $H^d$,
i.e we need to solve the eigenvalue problem  
$\Delta_{H^d} \alpha_n^{MN...}= - \lambda_n
\alpha_n^{MN..} $.
$\Delta_{H^d}$ acts differently on tensors of different ranks, 
however there are
common features for $\lambda$'s: they are all discrete, ordered
($\lambda_0\leq \lambda_1,...$), and bounded from below 
(being eigenvalues of a Laplacian on a compact manifold).
The zero mode of the Laplacian on compact space 
acting on a scalar field, 
$
\displaystyle
\Delta_{H^d}\phi(y)=\frac{1}{\sqrt{g}}\partial_i\left(
\sqrt{g}g^{ij}\partial_i\phi(y)
\right) $,
is
constant.
Therefore, the wave function of the {\it massless} graviton 
in $H^d$ is constant and the effective $M_{p}$ depends only on  
the volume factor. Using equation (\ref{v}) we can write:
\begin{equation}
M_{p}^2=M_*^{d+2}V
=M_*^{d+2}R^d \mbox{exp}((d-1)L/R)
\label{mass}
\end{equation}
By mildly tuning the diameter
$L\simeq 35 M_*^{-1}$ ($10^{-15}$ mm), 
the above equation (\ref{mass}) represents a good 
solution for the hierarchy problem, at least at the classical 
level. 


\subsection{Harmonic Spinors}


If one wants to end up with massless fermions
in 
$4$ dimensions after compactification, 
so that the standard model fermions get their masses 
through a Higgs mechanism, 
it is necessary
for the Dirac operator on the internal manifold, 
$\slash{D}$, to have
at least one zero mode.\footnote{Assuming that the compactified
manifold admits a spin structure.} 
As we pointed out in \cite{geo}, 
using earlier theorem by Lichnorowicz \cite{L}, 
positively curved compact manifolds do {\it not} admit 
harmonic spinors, 
while 
negatively curved compact do. This can be easily understood
looking at the eigenvalues of $\slash{D}^2$ (since 
$\mbox{ker}\slash{D}^2=\mbox{ker}\slash{D}$)
$$
\slash{D}^2=\nabla^*\nabla +\frac{1}{4}\cal R
$$
\noindent
Where $\cal R$ is the scalar curvature, and
$ \nabla^*\nabla $ is the connection Laplacian (a positive operator).

As can be easily verified, a manifold with a positive curvature,
like $S^2$ for example, does not admit harmonic (massless) spinors. 
Inorder to be able to get massless spinors on a sphere,  
it is necessary to couple the spinors to a magnetic
monopole.\footnote{I am grateful to Seif Randjbar-Daemi
for very useful discussions around this point.} 
In general, one has to do an extra labor in order
to get massless spinors by compactifying on positively curved 
manifolds, e.g. by 
twisting the Dirac operator of the internal space, or moding 
by its isometry group. On the other hand, one can generate 
naturally massless fermions by compactifying on manifolds of negative
curvature, like the CHMs.


\section{Some Phenomenological Disadvantages
of CHMs\label{dis}}


There is no analytical expression for the eigenvalues of 
the Laplacian on a generic compact manifold. Reliance on
mere dimensional analysis to set lower bounds 
may break down in some cases. 
In the work \cite{geo}
we used earlier results by \cite{yau}
and pointed out lower bounds on the first eigenvalue 
of the Laplacian acting on a scalar
on generic compact manifold $Y$:
\begin{equation}
\lambda_1 \geq \frac{\pi^2}{4L^2}-\mbox{max}\{-(d-1)K, 0\}
\label{lam}
\end{equation}
\noindent
where $\mbox{min}{\cal R}=(d-1)K$ (for 
constant curvature, ${\cal R}=(d-1)K$).
For ${\cal R}>0$, the fundamental parameter of the theory is the 
diameter, $L$ (the maximum distance on the manifold).
For ${\cal R}<0$ the curvature will also enter into the bounds.
In CHMs, within the approximation $L>R$ used
to solve
the hierarchy problem, there seem to 
be no lower bound on the first massive Kaluza--Klein 
graviton mode
from geometry, namely because the mass$^2$
will be bounded from below by a negative number 
$\displaystyle
m^2_{KK}\geq  
- \frac{1}{R^2}$.

Of course, this does not
mean that the values of $m^2_{KK}$ are not of the order TeV. The
fact that there exists a mass gap, makes this assumption
reasonable.   

We went further to discuss lower bounds on fermionic Kaluza-Klein 
modes 
for a generic compact $Y$. The first non-zero eigenvalue of 
$\slash{D}_Y$ on a compact space is bounded from below by 
\cite{hijazi}
\begin{equation}
m_e^2\geq \frac{d}{4(d-1)} \tau_1
\label{fermi}
\end{equation}
\noindent
Where $\tau_1$ is the first non-zero eigenvalue of the Yamabe 
operator
$\displaystyle {\cal L}=
\frac{4(d-1)}{d-2}\Delta_Y + {\cal R}$.
Using the bound (\ref{lam}) it is possible to see
that
for CHMs there are no geometric lower bounds on the 
fermions masses,  
within the approximation $L>R$.

\section{Conclusions}
A compact hyperbolic manifold, $H^d; d\neq 3$ ,
if proven to exist as a solution of the equations 
of motion for a particular geometry of the $4$ dimensional 
Universe, 
has some attractive features
which can be used in favor of the theories with large radii,
compactification: i) if the diameter of the CHM is slightly 
larger than the scale of its curvature, it is possible to have
a rather satisfactory {\it classical} solution for the hierarchy problem 
since only a very mild tuning for the parameters of the manifold
is required ii) CHMs admit harmonic spinors 
iii) all the Kaluza--Klein excitations of the metric are
heavy \cite{kn:m} $\geq 1/L$ (this issue was not discussed in this talk).
Unfortunately, as can be deduced from (\ref{lam}) and 
(\ref{fermi}), there are
no geometric lower bounds on the masses of Kaluza--Klein modes
of the gravity, 
fermions, and scalar fields. 
Hence relaxing the known bounds (like the upper bound on the 
reheating temperature and other astrophysical bounds) 
is not geometrically justified 
unless a further case-wise investigation is performed 
(see \cite{kn:m}
for the case of $H^2$).


\section*{Acknowledgments}


I would like to thank Avijit Mukherjee and Seif Randjbar-Daemi for 
useful and truly enlightening discussions.

\end{document}